 \documentstyle{article} 

\makeatletter
  \newlength\titlebox  
 \twocolumn 
 
\def\addcontentsline#1#2#3{} 
 
\def\maketitle{\par 
 \begingroup 
   \def\thefootnote{\fnsymbol{footnote}} 
   \def\@makefnmark{\hbox to 0pt{$^{\@thefnmark}$\hss}} 
   \twocolumn[\@maketitle] \@thanks 
 \endgroup 
 \setcounter{footnote}{0} 
 \let\maketitle\relax \let\@maketitle\relax 
 \gdef\@thanks{}\gdef\@author{}\gdef\@title{}\let\thanks\relax} 
\def\@maketitle{\vbox to \titlebox{\hsize\textwidth 
\linewidth\hsize \toptitlebar \centering 
 {\Large\bf \@title \par}  \bottomtitlebar \vskip 0.2in plus 1fil minus 0.1in 
 {\def\and{\unskip\enspace{\rm and}\enspace}%
  \def\And{\end{tabular}\hss \egroup \hskip 1in plus 2fil  
           \hbox to 0pt\bgroup\hss \begin{tabular}[t]{c}\bf}%
  \def\AND{\end{tabular}\hss\egroup \hfil\hfil\egroup 
         \vskip 0.25in plus 1fil minus 0.125in 
          \hbox to \linewidth\bgroup \hfil\hfil 
            \hbox to 0pt\bgroup\hss \begin{tabular}[t]{c}\bf} 
 \hbox to \linewidth\bgroup \hfil\hfil 
    \hbox to 0pt\bgroup\hss \begin{tabular}[t]{c}\bf\@author  
                         \end{tabular}\hss\egroup 
    \hfil\hfil\egroup} 
  \vskip 0.3in plus 2fil minus 0.1in 
}} 
\renewenvironment{abstract}{\centerline{\large\bf 
Abstract}\vspace{0.5ex}\begin{quote}}{\par\end{quote}\vskip 1ex} 
 
\def\section{\@startsection {section}{1}{\z@}{-2.0ex plus 
    -0.5ex minus -.2ex}{1.5ex plus 0.3ex minus .2ex}{\large\bf\raggedright}} 
\def\subsection{\@startsection{subsection}{2}{\z@}{-1.8ex plus 
    -0.5ex minus -.2ex}{0.8ex plus .2ex}{\normalsize\bf\raggedright}} 
\def\subsubsection{\@startsection{subsubsection}{3}{\z@}{-1.5ex plus 
   -0.5ex minus -.2ex}{0.5ex plus .2ex}{\normalsize\bf\raggedright}} 
\def\paragraph{\@startsection{paragraph}{4}{\z@}{1.5ex plus 
   0.5ex minus .2ex}{-1em}{\normalsize\bf}} 
\def\subparagraph{\@startsection{subparagraph}{5}{\z@}{1.5ex plus 
   0.5ex minus .2ex}{-1em}{\normalsize\bf}}

\skip\footins 9pt plus 4pt minus 2pt 
\def\footnoterule{\kern-3pt \hrule width 5pc \kern 2.6pt } 
\labelwidth\leftmargini\advance\labelwidth-\labelsep \labelsep 5pt 
 
\def\@listi{\leftmargin\leftmargini} 
\def\@listii{\leftmargin\leftmarginii 
   \labelwidth\leftmarginii\advance\labelwidth-\labelsep 
   \topsep 2pt plus 1pt minus 0.5pt 
   \parsep 1pt plus 0.5pt minus 0.5pt 
   \itemsep \parsep} 
\def\@listiii{\leftmargin\leftmarginiii 
    \labelwidth\leftmarginiii\advance\labelwidth-\labelsep 
    \topsep 1pt plus 0.5pt minus 0.5pt  
    \parsep \z@ \partopsep 0.5pt plus 0pt minus 0.5pt 
    \itemsep \topsep} 
\def\@listiv{\leftmargin\leftmarginiv 
     \labelwidth\leftmarginiv\advance\labelwidth-\labelsep} 
\def\@listv{\leftmargin\leftmarginv 
     \labelwidth\leftmarginv\advance\labelwidth-\labelsep} 
\def\@listvi{\leftmargin\leftmarginvi 
     \labelwidth\leftmarginvi\advance\labelwidth-\labelsep} 
 
\belowdisplayskip \abovedisplayskip 
\def\@normalsize{\@setsize\normalsize{11pt}\xpt\@xpt} 
\def\small{\@setsize\small{10pt}\ixpt\@ixpt} 
\def\footnotesize{\@setsize\footnotesize{10pt}\ixpt\@ixpt} 
\def\scriptsize{\@setsize\scriptsize{8pt}\viipt\@viipt} 
\def\tiny{\@setsize\tiny{7pt}\vipt\@vipt} 
\def\large{\@setsize\large{14pt}\xiipt\@xiipt} 
\def\Large{\@setsize\Large{16pt}\xivpt\@xivpt} 
\def\LARGE{\@setsize\LARGE{20pt}\xviipt\@xviipt} 
\def\huge{\@setsize\huge{23pt}\xxpt\@xxpt} 
\def\Huge{\@setsize\Huge{28pt}\xxvpt\@xxvpt} 
 
\def\toptitlebar{ 
\hrule height4pt 
\vskip .25in} 
 
\def\bottomtitlebar{ 
\vskip .25in 
\hrule height1pt 
\vskip .25in} 
 
\makeatother

 \def\itemhead#1{\item\emph{(#1)}\ \ }
 
 \title{What you always wanted to know about genetic algorithms\\ but were
afraid to hear%
 \\\vspace{.5ex}\small Revised (June 2000) from Lashon {\sc
Booker}, Stephanie {\sc Forrest}, Melanie {\sc Mitchell}, and Rick {\sc
Riolo} (ed.),\\ {\sl Festschrifts in honor of John Holland}, Center for the
Studies of Complex Systems, U.\ Michigan, Ann Arbor, June 1999.%
 }
 \author{{\bf Tommaso Toffoli} ({\tt tt@bu.edu})\\
 Electric and Computer Engineering\\Boston University, Boston, MA 02215} 
 
\begin{document} 
 \maketitle 
 \begin{abstract} 
In spite of their seemingly ``obvious'' virtues as a search strategy, genetic
algorithms have ended up playing only a modest role as design tools in
science and engineering. We review the reasons for this apparent failure,
and we suggest a more relaxed view of their utility.
 \end{abstract} 
 
\section{INTRODUCTION} 
 
``Adaptive evolution is the motor of biology. But its mechanisms are so
general that they should be effective in shaping artificial systems as
well!''  This was the manifesto John Holland read out at the beginning of the
70's. What many heard, or pretended to hear, was a message with more haste
and more hubris: ``Biological evolution works wonders. Let us apply its
methods to engineering and we'll work wonders ourselves!''

Thus, the \emph{genetic algorithm} (GA) soon found its way in the engineer's
toolbox, to be employed as a ready-to-use feedback loop for design
optimization (Shaffer 1999); the problem-specific part is supplied through a
``fitness function''. This combination works much like the generic
operational amplifier loop used in signal synthesis, where the engineer only
need drop in a problem-specific feedback element.

In thirty years, genetic algorithms have grown in expertise and
sophistication.  They have secured for themselves a place under the sun, but
in spite of their great promise they have not swept the engineering
world. Who should we complain with? And, then, about what?

\medskip

One may plead that today's genetic algorithms do not yet take advantage of
the evolutionary paradigm's full range of possibilities (cf.~Section
\ref{sect:wider}); that more storage and processing power are needed to do
full justice to the approach; that the software is still experimental and
cannot be readily used by the non-specialist. In other words, that better
performance is just a matter of time and patience (and, as usual, money).  We
shall argue that the core issue is a different one. Genetic algorithms are
\emph{not} on probation; they are already doing well today. They make a
superb tool for open-ended exploration. They may save conceptual labor (if
not necessarily material resources) in local optimization and dynamic
tracking tasks. But they fall in line with their competitors when it comes to
tasks of \emph{deliberate, long-haul design}---what is ordinarily meant by
engineering. Ours is not an empirical assessment: this failure is owed
to intrinsic factors. Namely, the very premises that make adaptive mechanims
work so well in the wild no longer hold in the engineer's studio.

\section{DELIBERATE VS ACCIDENTAL DESIGN}

We shall first present the issue is in terms of \emph{deliberate} vs
\emph{accidental} design.\footnote
 {We are not, of course, using `accidental' in a disparaging sense (Dawkins
1996).}
 This view, though correct, is not fully satisfactory. It allows one to
understand why genetic algorithms perform, under engineering constraints,
more poorly that one might hope, but does not really explain why the
performance gap between this mode of operation and that of unconstrained
evolution is so large. That will be done in Section \ref{sect:race}.

\medskip

Evolution indeed produces wonderful objects, but, like a stage magician, it
is in the enviable position not to have to state, ahead of time, what
precisely the next number will consist in. We may feel that, in designing a
cheetah, evolution had committed itself to optimizing for \emph{speed}---the
cheetah runs very fast indeed.  But the Darwininan ``contract'' did not
explicitly mention speed; it only stipulated differential advantage; this
might have been achieved by stealth instead of speed, or by a different
metabolism. The genuine objective function, ``long-term inclusive fitness'',
has many more peaks to offer than the more restrictive objective function
``speed''.  Only \emph{a posteriori} can one say that the beautiful design
which was delivered by evolution was ``for speed'': the next time we confront
evolution with a similar problem we may get a quite different solution. And
if we actually insist on putting ``speed'' in the contract
specifications---as when we do selective breeding---then we so much limit
evolution's options that a beautiful design may not come out at all.

In sum, to expect that evolution will perform on demand, on the shop floor,
stunts as amazing as those it performs in the wild is at the very least a
form of \emph{statistical fallacy!} In fact, the less we insist that
evolution read our wish list, the better will it be able to surprise us with
lovely presents. Unfortunately, though a resolve to be discreet in expressing
one's desires may make a good recipe for a lasting friendship, it would yield
a frustrating relationship with any engineering consulting firm: What is it
precisely that we want from them?

\section{THE CANONICAL SCENARIO}\label{sect:canonical}

Since the consolidation of the modern synthesis (cf.~Fisher 1930, Mayr
1963),\footnote
 {At least since 1868, we've had a plausible conceptual mechanism for
biological evolution, i.e., Darwin's theory of ``descent with
variation''; even though concrete implementation details were initially
missing, the theory did not depend much on them anyway. Nonetheless, Darwin's
theory remained exactly that, that is, just a \emph{theory} (and even one
that was not given particularly great attention to), until the burgeoning of
genetics.}
 evolutionary genetics has been using as a standard ``demonstration
set'' the nucleus of the eukaryotic cell---with a well-defined complement of
chromosomes undergoing meiosis and mitosis, mutation and crossover. Thus, the
storage and duplication of genetic data are kept in full view. On the other
hand, ontogenic development and interaction of the phenotype with its
environment do not explicitly appear on the scene; they are substituted for
by a narrator, the ``fitness function'', which just informs us of the drama's
final actuarial statistics---how many births and deaths.

This is the canonical scenario that Holland (1975) generalized to artificial
systems. By proceeding in this way he automatically guaranteed that genetic
algorithms \emph{can} work---since they include those of biology.  Moreover,
the scenario comes with an already well-developed body of theory connecting
gene mechanics to population statistics via an \emph{arbitrary}, externally
given fitness function. Thus, a generic genetic-algorithm ``driver'' can be
piggybacked on any combinatorial optimization problem. All you have to do is:
 \begin{itemize}
 \itemhead{Genome} Represent a search-space point by the string of its
coordinates in some appropriate reference frame.
 \itemhead{Population} Provide room for an adequately large number
of such strings.
 \itemhead{Fitness} Evaluate the given fitness function on each element
of the population; use the result to determine the relative size of its
offspring.
 \itemhead{Update step} Normalize (Hancock 1994), and replace the old
population with the new population.
 \end{itemize}
 Note that the GA driver does not need to know, for its proper
functioning, what the genome ``really means'': the responsibility for that
interpretation is carried solely by the external fitness function.  Thus, in
spite of their biological inspiration, genetic algorithms have no handicap in
dealing with problems of a nonbiological nature.

\medskip

Today, it is clear that the above scenario is only one of many in which
biological evolution can be found to act; though routine performances are
regularly shown on that stage, the most gripping dramas of evolution are
played more rarely and in less publicised venues (see Section
\ref{sect:wider}).  But, for the moment, let us stick with the canonical
scenario, and compare its uses by natural evolution with those by the
countless ``applications'' reported in a host of proceedings (see, for
instance, B\"ack 1997, Fogarty 1994, Higuchi 1997, Porto 1998).

The most striking difference is how directly, in the typical
artificial-system application, the phenotype is represented in the genotype.
In a traveling-salesman problem, the genotype may simply be an ordered list
of city coordinates (Valenzuela 1997); in a job shop scheduling problem, it
may be a list of the proposed starting times for the jobs to be performed
(Lin 1997); in an airplane design, a list of the numerical design parameters
to be varied (Rasheed 1997). Similarly, the typical fitness function involves
little more than metric or constraint information directly derivable from the
genome (distance between cities, conflict between jobs, etc.). The genetic
algorithm is thus asked to perform a task that not only is clerical, but that
cannot in principle be eased by any higher-level engineering insight---since
the problem has no hidden structure to be discovered. Other search approaches
like breadth-first or branch-and-bound may well give comparable (and
similarly unremarkable) results.  In these cases, the pedestrian performance
of the genetic algorithm directly reflects the pedestrian context in which it
was used.

\medskip

Let us turn to the use of the canonical scenario for more demanding
applications. Ideally, we would like the feedback loop from function to
genome, as established by the genetic algorithm, to ``servo'' a huge
collection of small clerical steps into an integrated engineering feat
involving many interlocking design aspects. After all, if the canonical
scenario has allowed nature to design the vertebrate eye, why shouldn't it
allow a GA to design a camera, given its functional specifications?

Imagine yourself trying to specify a fitness function that, starting with a
supply of assorted materials, will yield the design of an optical
camera. Your requirements are that
 \begin{enumerate}
 \item In combination with the generic GA driver, the fitness function shall
evolve the design for the camera in a reasonably short time. This requires
that selective pressure be maintained at a sustained level (and in the right
direction) throughout the whole process.  For a complex object like a camera
(lens, shutter, light-tightness of the enclosure, film transport) the only
way to insure this is for the fitness function to recognize possible
intermediate steps between the initial state and the completed camera, and
incrementally reward these steps in an appropriate order.\footnote
 {We have no qualms with arguments that \emph{individual} camera elements,
such as a lens, can rapidly evolve from a simple functional prescription
(Nilsson 1994).}
 \item On the other hand, the evolving system should not produce irrelevant
or confusing items (Rasheed 1997), such as something that looks like a camera
but doesn't quite work like one. In essence, we want to avoid the paradox of
the ``universal library'' (Lasswitz 1901)---which contains all possible books
and is therefore as useless as the empty library!  If the fitness function
can tell only at the last moment that the item produced after long labor is
not indeed a functional camera (and this would indeed be the case if the
fitness function only contained the bare functional specifications), then the
puzzle has to be scrambled and started all over again at every iteration.

 \end{enumerate}

In theory, the fitness function only needs to know what a camera should
\emph{do}. In practice, requirements 1 and 2 also ask this function to know
what a camera might look like and how it could be fabricated---but this is at
cross-purposes with the \emph{raison d'\^etre} of genetic algorithms, namely,
that
 \begin{enumerate}\setcounter{enumi}{2}
 \item Designing the fitness function should require much less knowledge and
engineering effort than directly designing the intended object in the first
place.
 \end{enumerate}
 It is this third requirement that the less pedestrian applications of
genetic algorithms in the canonical scenario tend to violate: a good design
may come out, but at the cost of a lot of problem-dependent knowledge that
one explicitly or implicitly supplied in order to set up an effective search
landscape.

\section{WIDER SCENARIOS}\label{sect:wider}

Perhaps, one might argue, requirements 1 and 2 of the previous section,
rather than having to be explicitly distilled by the engineer into the
fitness function, could be automatically catered to by a more flexible GA
driver. Why should a GA be obliged to stick to the canonical scenario when
natural evolution is allowed to run circles around it?

We barely need mention a host of possible variations on the
mutation/crossover theme, such as inversion, deletion, reduplication,
non-homologous crossover, dominance, polyploidy, etc., which are proposed
(though rarely used) in the GA literature. We also need barely mention other
external or internal factors (genetic drift, extreme geographical
segregation, geological catastrophes) which may momentarily let one escape
the tight control of the generate/select/regenerate loop. These are
relatively trivial enhancements to the canonical scenario, and may be readily
incorporated into it.

A more significant hint as to how genetic algorithms could be
enhanced is provided by the frequency with which
a \emph{symbiosis} ``operator'' has intervened in
the major evolutionary inventions (Margulis 91). Among the most familiar
forms of symbiosis one may list the intimate compact between two species, as
exemplified by lichens, the vital interdependence between individuals of a
bee colony, and the exacting social contract\footnote
 {For example, the blood's red cells are asked to forfeit reproduction (they
don't have a nucleus) for the sake of gas transport efficiency, and sperms
agree to leave behind their mitochondria on entering the egg.}
 between the cells of a multicellular organism. In turn, the ``legal machinery''
that makes this contract operative is embodied by a more subtle form of
symbiosis, that is, the complex relationship between genome and
soma (Dawkins 1990, Ridley 1995).

Less widely appreciated is the symbiosis between different bacteria, which,
established early in the history of life, soon gave rise to the eukaryotic
cell (Margulis 1995). The establishment of a symbiosis is an inversion of the
mechanism which is central to so many artificial intelligence models. Instead
of a main goal recursively generating a family of subgoals, here we have
instead a set of independently born goals that eventually and accidentally
accrete into a supergoal, and may do so recursively.  With reference to the
discussion of the previous section, when the fitness function of the
supergoal needs to be optimized, the symbionts have already been optimized
according to \emph{their own} fitness functions---so that only fine tuning of
the overall function is needed.  Thus, one gets most of the benefits of a
variable or hierarchical fitness function without the costs of having to
design it (cf.~ Smith 1776); the catch, of course, is that one has no idea
where the whole process is headed (Schmookler 1993).

To proceed even further on this theme, cultural evolution, with novel
reproductive mechanisms (Dawkins 1990), is superposed on ordinary genetic
evolution and maintains an intimate relationship with it (Plotkin 1993).
Finally, sociality and culture come together in large human
organizations. Some of these organizations may enjoy enough distinctness,
permanence, and specific reactivity to qualify as collective organisms with
an individuality of their own, opening up still further evolutionary
possibilities.\footnote
 {Note that a large fraction of  today's software engineering effort
is aimed at providing these organization with a ``nervous system''
of their own.}

\medskip

On the artificial front, cellular automata and swarm systems explore
collective behavior at different levels of aggregation.  Synthetic
evolutionary system such as Tierra and its descendants (Ray 1994) have been
quite successful at inventing from scratch forms of collaboration,
antagonism, symbiosis, and parasitism. In fact, no serious artificial life
projects can do without provisions for open-ended, hierarchical evolution.

\medskip

Would genetic algorithms be more effective engineering tools if their
repertoire included such extended forms of evolution, or if, like Tierra,
they were provided with the means to ``roll their own''? In fact, multi-level
genetic algorithms are occasionally found. Simulated annealing with
``designer's'' temperature schedules provides a systematic way to designate a
particular aggregation level as the temporary focus of evolutionary activity
(Ingber 1992). Even more integration between levels is envisioned by Sanchez
(1997). However, it is doubtful that such measures, drastic as they are,
could make much of a difference with respect to the issue discussed here,
namely, the performance gap between the exploratory and the engineering mode
of genetic algorithms, as I'll explain in the next section.

 \section{ON A RACE WITH ONESELF}\label{sect:race}

With a \emph{fixed} fitness function, genetic algorithms can only do so
much---but, if Holland's arguments about the equivalence between natural and
artificial adaptation are correct, this must also be the case for natural
evolution.  In fact, hardly ever does one catch the latter doing other than
routine maintenance and minor parameter tracking: the average duration of a
species is a few million years, and during this time the phenotype hardly
shows any changes (Eldredge and Gould 1972).

As we saw in Section \ref{sect:canonical}, if one were allowed to put into
the fitness function a sequence of incrementally graded intermediate goals
(``If in state $S_1$ please evolve to $S_2$; if in $S_2$, try to get to
$S_3$;'' and so forth), or, equivalently, if one were allowed to use a
variable fitness function that swept through those states, then engineering
design via a GA would be a breeze.  But point 3 stipulated that the fitness
function should, in essence, consist of no more that the functional
specifications.\footnote
 {Consider the \emph{indicator function}---which returns a 1 if the design
meets the functional specifications and 0 otherwise. The typical
fitness function is just a smoothed-out version of the indicator function.
In principle, any ``massaging'' of the indicator function is allowed,
but the massaging rule should be part and parcel of the GA package, not
a problem-dependent transformation.}
 In other words, since in the engineering design mode the target is
fixed---by contract, as it were---and the fitness function encodes the target
in a fixed way, then it is in the very nature of this design mode that the
fitness function itself should be fixed.

\medskip

In natural adaptation, however, the fitness function does not stay fixed
\emph{all the time}. If the environment goes through significant
alterations, so does the fitness function. Here we distinguish three cases.
 \begin{enumerate}
 \item If environmental changes take place very slowly, then species easily
adjust to them; but, precisely because the changes are so small, no
evolutionary feats of great consequence may be expected. In this case, the
achievements of natural evolution certainly do not threaten the engineering
studio's pride.
 \item If the change is sudden and large, then natural adaptation is unable
to respond and extinction occurs. Again, nothing here for evolution to brag
about.
 \item If the rate of change of the environment is large but still within the
tracking range of the adaptive mechanism, then indeed evolution will proceed
at a fast rate. But how often will environmental change happen to meet these
requirements? If it slows down, we go back to case 1. If it speeds up, we
fall into case 2. If it goes back-and-forth, the instantaneous rate of
evolution will---it is true---be proportional to that of these changes, but
the long-term rate will only be proportional to its \emph{square
root}---since we have a diffusive process.
 \end{enumerate}
 Thus, one might conclude that the long-term average of a natural system's
evolution rate will never amount to much. But this conclusion is spoiled by a
special situation that occurs within case 3, and which is conceivably
responsible for the bulk of evolutionary change. This is when the significant
environment of an individual is represented not only by an unresponsive
geological/astronomical background, but also by the other individuals of the
same species and by a few other species that tightly interact with it.  In
these circumstances, as soon as a species changes so does its fitness
function, creating an immediate feedback. The positive component of
this feedback may put the species in a sustained race with itself: As soon as a
tree discovers how to get taller than the others and capture more sunlight,
its success, i.e., enhanced reproduction, makes the whole population taller;
then, to stay ahead, a tree has to get even taller.

Since here the bar is raised by the organism itself, not by external,
unrelated forces, there is no risk that it will be raised too high too soon,
as in case 2 above. That is, positive feedback is automatically limited by a
negative feedback component that adjust the rate of change of a species, seen
as environment, to that sustainable by the species seen as an adaptive
system: If a crab's claws become too successful at crushing oysters, and thus
at \emph{acquiring} food, then oysters disappears fast and the crab becomes
by the same token less successful at \emph{finding} food; in this sense, claw
strength is self-limiting.

Such a controlled arms race is precisely the regime where evolution runs
close to its maximum possible speed (Stanley 1981, Ridley 1995). Even
relatively brief spells of it can drastically raise the overall long-term
rate (since, as we've seen, evolution's background rate is close to zero). We
may call this kind of design \emph{accidental}, but we have to acknowledge it
is incomparably faster than the typical \emph{deliberate}, fixed-target
design characteristic of the engineering mode.

 \section{CONCLUSIONS}

The evolutionary performance of natural life may constitute a reasonable
benchmark for artificial life, but is not a meaningful term of comparison for
genetic algorithms as used in the design engineer's studio. A studio is asked
to produce designs on contract; evolution volunteers, when and where it
pleases, objects of its own choice that ``look as'' they were designed---what
Dawkins (1996) calls \emph{designoids}.

 We pay the engineering studio not for the \emph{information} its product
contains, but for the \emph{correlation} that the design process guarantees
between the product and our stated needs (cf.\ Plotkin 1993, Adami
1996).\footnote
 {In an analogous fashion, criminal law metes out punishment on the basis of
intention rather than outcome, as when it distinguishes between murder and
accidental manslaughter.}
 Genetic algorithms shouldn't feel slighted if the engineer, having to give
these needs immediate priority, is forced to use a powerful tool in a 
much more conservative fashion than natural evolution.

 \section*{REFERENCES}
 \def\bibitem#1{\par}
 \let \bib\bibitem
 \bib{adami}{\sc Adami}, Chris, and Nicolas {\sc Cerf}, ``Complexity,
computation, and measurement'', in {\sc Toffoli} et al. (eds.), {\sl Physics
and Computation}, New England Complex Systems Institute (1996), 7--11; a
revised version to appear in {\sl Physica D}.
 \bib{back}{\sc B\"ack}, Thomas (ed.), {\sl Genetic Algorithms}, Morgan
Kaufmann (1997).
 \bib{dawkins:gene}{\sc Dawkins}, Richard, {\sl The Selfish Gene}, Oxford
Univ.\ Press (1990).
 \bib{dawkins:pheno}{\sc Dawkins}, Richard, {\sl The Extended Phenotype: The
Long Reach of the Gene}, Oxford Univ.\ Press (1990).
 \bib{dawkins:mount}{\sc Dawkins}, Richard, {\sl Climbing Mount Improbable},
Norton (1996).
 \bib{punctuated}{\sc Eldredge}, Niles, and Stephen {\sc Gould},
``Punctuated equilibria: An alternative to phyletic gradualism'', in
T. {\sc Schopf} (ed.), {\sl Models in Paleobiology}, Freeman (1972), 82--115.
 \bib{fisher}{\sc Fisher}, R. A., {\sl The Genetical Theory of Natural
Selection}, Clarendon Press (1930).
 \bib{hancock}{\sc Hancock}, Peter, ``An empirical comparison of selection
methods in evolutionary algorithms'', {\sl Evolutionary Computing}, Springer--Verlag (1994), 80--94.
 \bib{higuchi}{\sc Higuchi}, Tetsuya, Iwata {\sc Masaya}, and Weixin {\sc Liu}
(eds.), {\sl Evolvable Systems: From Biology to Hardware}, Springer (1997).
 \bib{holland}{\sc Holland}, John, {\sl Adaptation in Natural and Artificial
Systems}, University of Michigan (1975), reprinted with revision by MIT Press
(1992).
 \bib{ingber}{\sc Ingber}, L., and B. Rosen, ``Adaptive simulated annealing
wins'', {\sl Math.\ Comp.\ Modeling} {\bf 16} (1992), 87--100.
 \bib{jaynes:principle}{\sc Jaynes}, Edwin T., ``Information theory and
statistical mechanics'', {\sl Phys.\ Rev.} {\bf 106} (1957), 620--630;
{\bf 108} (1957), 171--190.
 \bib{jaynes:minimum-product}{\sc Jaynes}, Edwin T., ``The minimum entropy
production principle'', {\sl Ann.\ Rev.\ Phys.\ Chem.} {\bf 31} (1980),
579--601.
 \bib{lasswitz}{\sc Lasswitz}, Kurd, ``The universal library'' (1901),
translated from the German by Willy Ley and republished in Clinton {\sc
Fadiman} (ed.), {\sl Fantasia Mathematica}, Simon and Schuster (1958),
237--243.
 \bib{lin}{\sc Lin}, Shyh-Chang, Eric {\sc Goodman}, and William {\sc Punch},
``A genetic algorithm approach to dynamic job shop scheduling problems'', in
Thomas {\sc B\"ack} (ed.), {\sl Genetic Algorithms}, Morgan Kaufmann (1997),
481--488.
 \bib{margulis91}{\sc Margulis}, Lynn, and Rene {\sc Fester} (eds.), {\sl
Symbiosis As a Source of Evolutionary Innovation: Speciation and
Morphogenesis}, MIT Press (1991). 
 \bib{margulis95}{\sc Margulis}, Lynn, {\sl Symbiosis in Cell Evolution:
Microbial Communities in the Archean and Proterozoic Eons}, W H Freeman
(1995)
 \bib{mayr}{\sc Mayr}, Ernst, {\sl Animal Species and Evolution}, Harvard
U. Press (1963).
 \bib{nilsson}{\sc Nilsson}, D.--E., and S. {\sc Pelger}, ``A pessimistic
estimate of the time required for the eye to evolve'', {\sl Proc.\ R.\ Soc.\
London B} {\bf 256}, 53--58.
 \bib{plotkin}{\sc Plotkin}, Henry, {\sl Darwin Machines and the Nature of
Knowledge}, Harvard Univ. Press (1993), xviii+269 pp.
 \bib{prigogine}{\sc Prigogine}, Ilya, {\sl Order out of Chaos}, Bantam
(1986).
 \bib{rasheed}{\sc Rasheed}, Khaled, and Hayim {\sc Hirsh}, ``Using
case-based learning to improve genetic-algorithm-based design optimization'',
in Thomas {\sc B\"ack} (ed.), {\sl Genetic Algorithms}, Morgan Kaufmann
(1997), 513--520.
 \bib{ray}{\sc Ray}, Tom, ``Evolution, complexity, entropy, and artificial
reality'', {\sl Physica D} {\bf 75} (1994), 239--263.
 \bib{ridley}{\sc Ridley}, Matt, {\sl The Red Queen: Sex and the Evolution of
Human Nature}, Penguin USA (1995), ix+405 pp.
 \bib{sanchez}{\sc Sanchez}, Eduardo, et al., ``Philogeny, ontogeny, and
epigenesis: three sources of biological inspiration for softening hardware'',
in Tetsuya {\sc Higuchi} et al. (eds.), {\sl Evolvable Systems: From Biology
to Hardware}, Springer (1997), 35--54.
 \bib{schaffer}{\sc Schaffer}, J. D., ``Practical guide to genetic
algorithms'',
{\tt chem1.nrl.}\linebreak[0]{\tt navy.mil/}\linebreak[0]{\tt\char"7Eshaffer/}
\linebreak[0]{\tt practga.}\linebreak[0]{\tt html}
(1999).
 \bib{schmookler}{\sc Schmookler}, Andrew, {\sl The Illusion of Choice: How
the Market Economy Shapes Our Destiny}, State University of New York (1993).
 \bib{smith:nations} {\sc Smith}, Adam, {\sl The Wealth of Nations: An Inquiry
into the Nature and Causes}, (1776); reprinted, Modern Library (1994).
 \bib{stanley}{\sc Stanley}, Steven, {\sl The Evolutionary Timetable}, Harper
(1981).
 \bib{valenzuela}{\sc Valenzuela}, Christine, and L. P. {\sc Williams},
``Improving heuristic algorithms for the travelling salesman problem by using
a genetic algorithm to perturb cities'', in Thomas {\sc B\"ack} (ed.), {\sl
Genetic Algorithms}, Morgan Kaufmann (1997), 458--464.

\end{document}